\documentclass[twocolumn,showpacs,prd,floatfix,axodraw]{revtex4}
\usepackage{mathrsfs}
\usepackage{graphicx,,booktabs,bm}
\usepackage{overpic}
\usepackage{color}
\usepackage{amssymb}

\usepackage{mathrsfs,bm,amsmath,amssymb}
\usepackage{longtable,lscape}
\usepackage{txfonts}
\usepackage{amssymb}
\usepackage{indentfirst}
\usepackage{graphicx,,booktabs}
\usepackage{multirow}
\usepackage{color}
\usepackage{amssymb}

\definecolor{cover}{rgb}{0.77,0.87,0.88}
\definecolor{blueone}{rgb}{0.1,0.1,.7}
\definecolor{citec}{rgb}{0.14,0.47,0.09}
\definecolor{two}{rgb}{0.0,0.5,0.}
\definecolor{three}{rgb}{.5,.1,0.15}
\usepackage[bookmarks=true,bookmarksopen=false,plainpages=false,breaklinks=true,
   bookmarksnumbered=true,hypertexnames=false,
   filecolor=blue,urlcolor=three,menucolor=three,
   linkcolor=three,citecolor=blueone, colorlinks,
   anchorcolor=blue,runcolor=pink,frenchlinks=red
   pdfstartview=FitH,pdftitle=title,%
   pdfauthor=author]{hyperref}

\begin{document}
\title{$\Xi(2030)$ and $\Xi(2120)$ as  $\bar{K}^{*}\Sigma$  molecular states}

\author{Yin Huang$^1$}
\email{huangy2017@buaa.edu.cn}
\author{Rong Wang$^{2}$}
\author{Jun He$^{3}$}
\email{Corresponding author: junhe@njnu.edu.cn}
\author{Ju-Jun Xie$^{4}$}
\author{Lisheng Geng$^1$}
\email{Corresponding author: lisheng.geng@buaa.edu.cn}

\affiliation{$^1$School of Physics and Nuclear Energy Engineering,
International Research Center for Nuclei and Particles in the Cosmos and
Beijing Key Laboratory of Advanced Nuclear Materials and Physics,
Beihang University, Beijing 100191, China\\
$^2$Institut de Physique Nucl\'eaire,
CNRS-IN2P3, Univ. Paris-Sud, Universit\'e Paris-Saclay,
91406 Orsay Cedex, France\\
$^3$Department of Physics and Institute of Theoretical Physics, Nanjing Normal University,
Nanjing 210097, People¡¯s Republic of China\\
$^4$Institute of Modern Physics, Chinese Academy of Sciences, Lanzhou 730000, China}

\date{\today}
\begin{abstract}
In this work, the  molecular states  from the $\bar{K}^{*}\Sigma$ interaction are studied in a quasipotential Bethe-Salpeter equation approach with the one-boson-exchange potentials. We consider the exchanges of vector ($\rho,\omega$ and $\phi$)  mesons
and pseudoscalar ($\pi$ and $\eta$) mesons  to describe the $\bar{K}^{*}\Sigma$ interaction with
 the coupling constants determined by the SU(3) symmetry,   The poles of the scattering
amplitude are searched for to find the molecular states dynamically generated from the $\bar{K}^{*}\Sigma$ interaction.   A bound state with quantum number $I(J^{p})=1/2(5/2^{+})$
is produced, which  can be associated to the $\Xi(2030)$ baryon.    Other  bound states with $J\le5/2$ are also searched for  and it is found that there exist bound states with quantum numbers $I(J^P)=1/2(1/2^-)$, $1/2(3/2^{-})$, and $1/2(3/2^+)$.   The $\Xi(2120)$ may be a candidate for one of these three bound states.   In addition, we also find two bound states with isospin $3/2$ and spin parities $1/2^-$ and $1/2^+$.
\end{abstract}


\maketitle
\section{Introduction}\label{sec1}
The observation of the hidden-charmed pentaquark states, $P_c(4450)$ and $P_c(4380)$, by the LHCb collaboration has reignited interests in  exotic baryons both
theoretically and experimentally~\cite{Aaij:2015tga}.  Many explicit models have been proposed to understand these exotic baryons~\cite{Chen:2016qju}.
Considering the progress in the heavy flavor sector, it is interesting to study possible exotic baryons in the light flavor sector.

In fact,  even before the quark model was proposed by Gell-Mann and Zweig, the  $\Lambda(1405)$ had been suggested as a bound state from the $\bar{K}N$ interaction, $i.\ e.$, a hadronic molecular state in modern terminology~\cite{Dalitz:1959dn}.  After its introduction, the quark model has been remarkably  successful in explaining most of the known hadrons as either $q\bar{q}$ mesons or $qqq$ baryons~\cite{Isgur:1978xj,Capstick:1986bm}.  However,  there exist some baryons that can not be easily fitted into
 the $qqq$ picture.  For instance, in constituent quark models, the mass of the strange quark is about 50\% heavier than the mass of the $u/d$ quarks.  This makes it difficult to understand why the mass of the $\Lambda(1405)$ is much lower than the mass of the $N(1535)$.  It is also confusing to see that the $N(1440)$ as a $N$=2 baryon is much lighter than the nucleon resonance $N(1535)$ with $N=1$, where $N$ is the main quantum number.

These issues and others, such as the strange magnetic momentum problem,  inspired Zou and his collaborators to propose that there should exist considerable five-quark components in  the nucleon and its resonances~\cite{Zou:2005xy,An:2005cj,Riska:2005bh}.  The mass inversion problem could be understood if  there are
significant five-quark ($uuds\bar{s}$) components in the $N(1535)$~\cite{Zou:2010tc,Zou:2009zz}, and the five-quark
configurations also provided a natural explanation for  its large couplings  to the strange $K\Lambda,K\Sigma,N\eta^{'}$, and  $N\phi$ channels~\cite{Liu:2005pm,Xie:2007qt}. This idea was extended to the charm sector and the hidden-charmed baryons were predicted in Refs.~\cite{Wu:2010jy,Yang:2011wz,Yuan:2012wz}. The existence of such exotic baryons was later confirmed by the LHCb observation, which has led to a revival of pentaquark studys~\cite{Aaij:2015tga}.
It is the purpose of the present work  to study possible exotic baryons in the light flavor sector.

Though there still exist some debates, it seems more natural to interpret the  $\Lambda(1405)$ as a molecular state  of
$\bar{K}N$ and its coupled channel~\cite{Patrignani:2016xqp,Oset:1997it,Oller:2000fj,Hyodo:2011ur}, which was also supported by the lattice-QCD simulations~\cite{Nemoto:2003ft,Hall:2014uca}. Recently,  inspired by the  LHCb observation of $P_c(4380)$ and $P_c(4450)$ and their  interpretation as $\Sigma^*_c\bar{D}$ and $\Sigma_c\bar{D}^*$~\cite{He:2015cea,He:2016pfa,Roca:2016tdh,Roca:2015dva} molecules, two nucleon resonances with a mass about 2.0 GeV, the $N(1875)$ and the $N(2100)$, were also interpreted as hadronic molecular states from the $\Sigma^{*}K$ and $\Sigma{}K^{*}$ interactions, respectively~\cite{He:2017aps,He:2015yva}.  Such an assignment is also supported by a recent calculation in the constituent quark model~\cite{Huang:2018ehi}.  The $\Sigma^{(*)}K^{(*)}$ system has strangeness zero and behaves as a nucleon resonance.   If we change the $K^{(*)}$ to a $\bar{K}^{(*)}$,  the system becomes double-strange, and can be related to a $\Xi$ baryon.  Different from the very recent observation of the double-charmed baryon $\Xi_{cc}$, several double-strange $\Xi$ baryons have been listed in the review of particle data group (PDG)~\cite{Patrignani:2016xqp}. Considering the interpretations of the $\Lambda(1405)$ as a $\bar{K} N$ bound  state and the $N(1875)$ and $N(2100)$ as $\Sigma^* K$ and $\Sigma K^*$ molecular states, it is interesting to study whether $\Sigma^{(*)}$ and $\bar{K}^{(*)}$ can be bound together to form a hadronic molecular state corresponding to
a $\Xi$ baryon.

At present, there are eleven $\Xi$ baryons listed in the review of PDG~\cite{Patrignani:2016xqp}.
The ground-state octet and decuplet baryons, the $\Xi(1320)$ and the $\Xi(1530)$, are well established with four-star ratings and can  be easily
fitted into the conventional quark model.  For the states $\Xi(1690)$, $\Xi(1820)$, $\Xi(2030)$ and $\Xi(2120)$,  there exist many different  interpretations, such  as $qqq$ states, molecular systems, etc~\cite{Sekihara:2015qqa,Kolomeitsev:2003kt,Sarkar:2004jh,Chao:1980em,Pervin:2007wa,Xiao:2013xi}.  In this work, we study  the   $\bar{K}^*\Sigma$ interaction and try to understand the double-strange baryons, the $\Xi(2030)$ and the $\Xi(2120)$, which have masses close to the $\bar{K}^*\Sigma$  threshold.

The  $\Xi(2030)$ is a three-star state and has a mass of $2025\pm5$ MeV and a width of $20^{+15}_{-5}$ MeV~\cite{Patrignani:2016xqp}. An early experimental analysis ~\cite{Hemingway:1977uw} suggested that the spin of the $\Xi(2030)$ should be $J\geq{}5/2$.  Before the experimental observation of the $\Xi(2030)$,   Samios et al. predicted that according to the SU(3) flavor symmetry  the $\Xi$(2030) is most likely the partner of the $N(1680)$, $\Lambda(1820)$ and $\Sigma(1915)$ with $J^P=5/2^{+}$~\cite{Samios:1974tw}.  The constituent quark model calculations in Refs.~\cite{Chao:1980em,Pervin:2007wa}
indicated that the $\Xi(2030)$ might be a candidate for the second $J^{P}=5/2^{+}$ state or $J^{P}=7/2^{+}$ state.
However, the strong decay analysis based on the experimental measurements disfavors the assignment of  the $\Xi(2030)$ as a member of the $5/2^+$ octet~\cite{PavonValderrama:2011gp} .  Also, the strong decay  analysis in the chiral quark model~\cite{Xiao:2013xi}
concludes that the $\Xi(2030)$ could not be assigned as any spin-parity $J^P=7/2^{+}$ states or  pure $J^P=5/2^{+}$ staes.  It seems
to favor the $J^P=3/2^{+}$ assignment. However, this conflicts with the analysis of the data~\cite{Hemingway:1977uw}. Generally speaking, the $\Xi(2030)$ can not
easily be understood as a conventional $qqq$ state.

 Compared with the $\Xi(2030)$, the experimental information on  the one-star state $\Xi(2120)$  is scarce~\cite{Patrignani:2016xqp}:
 both the spin parity  and the width is not known experimentally. The $\Xi(2120)$ was first observed in the $\bar{K}\Lambda$ invariant mass spectrum by  the Amsterdam-CERN-Nijmegen-Oxford Collaboration~\cite{Gay:1976sr}, and later confirmed  by the French-Soviet and CERN-Soviet Collaboration~\cite{Vorobev:1979as} in the 1970's, where a mass of about 2120 MeV and a width of about 20 MeV were suggested by those observations with poor statistics.   There exist a few theoretical  studies about the nature of  the $\Xi(2120)$.  A study in the chiral unitary approach suggested that a pole around 2100 MeV can be produced from the interaction between pseudoscalar/vector mesons and baryons with $J^P=1/2^-$ and $3/2^-$, which can be associated to the $\Xi(2120)$~\cite{Oset:2009vf,Gamermann:2011mq}.  In Ref.~\cite{Khemchandani:2016ftn} the $I(J^{P})=1/2(3/2^{-})$ state
located at $2046-i8.2$ MeV is identified as a meson-baryon molecule that can be associated to the $\Xi(2120)$.  Meanwhile, it is claimed that the $\Xi(2120)$ have
a big $\bar{K}^{*}\Sigma$ component.

From the above discussion, it is clear that the $\Xi(2030)$ can not be easily understood as a $qqq$ state. The mass of  the $\Xi(2030)$ is close to the $\bar{K}\Sigma^{*}$
threshold, which encourages us to interpret $\Xi(2030)$ as a bound state of $\bar{K}^{*}\Sigma$.  The $\Xi(2120)$ was also interpreted as a molecular state in the literature~\cite{Oset:2009vf,Gamermann:2011mq,Khemchandani:2016ftn}.  It is natural to interpret the $\Xi(2030)$ and $\Xi(2120)$ baryons as two bound states from the $\bar{K}^*\Sigma$ interaction with different quantum numbers.  It is analogous to the interpretations of the LHCb pentaquark states,  $P_c(4450)$ and $P_c(4380)$,
as bound states of $\bar{D}^*\Sigma$  with spin parities $5/2^+$ and $3/2^-$~\cite{He:2016pfa}.  With such an assumption, the $\Xi(2030)$ is at least a $P$-wave bound state. According to  Ref.~\cite{He:2016pfa}, a molecualr state in $P$ wave can still be produced and observed in experiment. In this work, the $\bar{K}^{*}\Sigma$ interaction will be studied in a quasipotential Bethe-Salpeter equation and the bound states, especially those relevant to the $\Xi(2030)$ and $\Xi(2120)$ will be searched for to investigate the possibility of  interpreting both $\Xi(2030)$ and $\Xi(2120)$ as hadronic molecular states.

This paper is organized as follows. The theoretical formalism is explained in Sec.~\ref{Sec: formulism}.  In Sec.~\ref{Sec: results}, the molecular states
of $\bar{K}^{*}\Sigma$ are searched for by looking for poles of the scattering amplitude on the complex plane. In the last section, we
provide a short summary and outlook.

\section{Theoretical framework }\label{Sec: formulism}
\begin{figure}[h!]
\begin{center}
\includegraphics[scale=0.65]{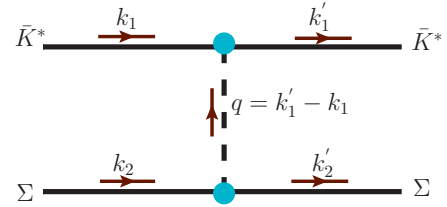}
\caption{Feynman diagram for the $\bar{K}^{*}\Sigma\to\bar{K}^{*}\Sigma$ interaction.  We also show the definition of the kinematics ($k_1,k_1^{'},k_2,k_2^{'}$) that we use to
calculate the potential kernel.}
\label{feynman-12}
\end{center}
\end{figure}

In this section,  we construct the  potential kernel of the  $\bar{K}^{*}\Sigma$  interaction described by the one-meson-exchange model~\cite{Machleidt:1987hj}.  For the system studied here, the pseudoscalar $\pi$ and $\eta$ mesons and vector $\rho$, $\omega$ and $\phi$ mesons are considered.  The  tree level Feynman diagram is illustrated in Fig.~\ref{feynman-12}.

The effective Lagrangian terms required for the relevant interaction vertices have the following form~\cite{He:2017aps}:
\begin{align}
&{\cal{L}}_{\bar{K}^{*}\bar{K}^{*}V}=i\frac{g_{\bar{K}^{*}\bar{K}^{*}V}}{2}(\bar{K}^{*\mu}V_{\mu\nu}\bar{K}^{*\nu}+\bar{K}^{*\mu\nu}V_{\mu}\bar{K}^{*\nu}+\bar{K}^{*\mu}V_{\nu}\bar{K}^{*\nu\mu})\\
&{\cal{L}}_{\bar{K}^{*}\bar{K}^{*}P}=g_{\bar{K}^{*}\bar{K}^{*}P}\epsilon^{\mu\nu\alpha\beta}\partial_{\mu}\bar{K}^{*}_{\nu}\partial_{\alpha}P\bar{K}^{*}_{\beta},\\
&{\cal{L}}_{\Sigma\Sigma{}V}=-g_{\Sigma\Sigma{}V}\bar{\Sigma}[\gamma^{\nu}-\frac{\kappa_{\Sigma\Sigma{}V}}{2m_{\Sigma}}\sigma^{\nu\alpha}\partial_{\alpha}]V_{\nu}\Sigma.\\
&{\cal{L}}_{\Sigma\Sigma{}P}=-\frac{g_{\Sigma\Sigma{}P}}{m_{\pi}}\bar{\Sigma}\gamma^{5}\gamma^{\mu}\partial_{\mu}P\Sigma,
\end{align}
where $V$ denotes $\rho,\omega,\phi$ and $P$ denotes $\pi,\eta$.  For an isovector $V$ and $P$, $V\to{}\vec{V}\cdot\tau$ and $P\to{}\vec{P}\cdot\tau$ if the vertices are $\bar{K}^{*}\bar{K}^{*}V$ and $\bar{K}^{*}\bar{K}^{*}P$, respectively.   If the vertices are ${\Sigma\Sigma{}V}$ and ${\Sigma\Sigma{}P}$ instead, then $V\to{}\vec{V}\cdot{}\vec{T}$ and $P\to{}\vec{P}\cdot{}\vec{T}$,
where $\vec{T}$ is the isospin transition operator~\cite{Matsuyama:2006rp}.  The $m_{\pi}$ and $m_{\Sigma}$ are the mass of $\pi$ and the nucleon, respectively.  The $K^{*}$ isodoublets are defined as
\begin{equation}
K^{*}=
\left(
  \begin{array}{c}
    K^{*+} \\
    K^{*0} \\
  \end{array}
\right);~~~~~
\bar{K}^{*}=
\left(
  \begin{array}{cc}
    K^{*-}& \bar{K}^{*0} \\
  \end{array}
\right).
\end{equation}
The $\vec{\pi},\vec{\rho}$ and $\vec{\Sigma}$ represents the $\pi,\rho$ and $\Sigma$ triplets
\begin{equation}
\vec{\pi}=
\left(
  \begin{array}{ccc}
    \pi^{+}\\
   \pi^{0} \\
   \pi^{-} \\
  \end{array}
\right);
\vec{\rho}=
\left(
  \begin{array}{ccc}
    \rho^{+}\\
    \rho^{0} \\
   \rho^{-} \\
  \end{array}
\right);
\vec{\Sigma}=
\left(
  \begin{array}{ccc}
    \Sigma^{+}\\
    \Sigma^{0} \\
   \Sigma^{-} \\
  \end{array}
\right).
\end{equation}
The field-strength tensor are defined as $V_{\mu\nu}=\partial_{\mu}V_{\nu}-\partial_{\nu}V_{\mu}$ and $\bar{K}_{\mu\nu}=\partial_{\mu}\bar{K}_{\nu}-\partial_{\nu}\bar{K}_{\mu}$.
The $\vec{\tau}$ is the Pauli matrix, while $\epsilon^{\mu\nu\alpha\beta}$ is the Levi-Civit$\grave{a}$ tensor.  In the above
Lagrangian terms,  $\sigma_{\mu\nu}=\frac{i}{2}(\gamma_{\mu}\gamma_{\nu}-\gamma_{\nu}\gamma_{\mu})$.

According to the SU(3) symmetry, the coupling constants have the following relation~\cite{deSwart:1963pdg},
\begin{align}
&g_{\bar{K}^{*}\bar{K}^{*}\rho}=g_{\bar{K}^{*}\bar{K}^{*}\omega}=\frac{g_{\bar{K}^{*}\bar{K}^{*}\phi}}{\sqrt{2}}=\frac{g_{\rho\rho\rho}}{2},\\
&g_{\Sigma\Sigma\rho}=g_{\Sigma\Sigma\omega}=2\alpha_{NN\rho}g_{NN\rho},\\
&g_{\Sigma\Sigma\phi}=-\sqrt{2}(2\alpha_{NN\rho}-1)g_{NN\rho},\\
&g_{\bar{K}^{*}\bar{K}^{*}\pi}=\frac{g_{\bar{K}^{*}\bar{K}^{*}\eta}}{\sqrt{3}}=\frac{g_{\omega\rho\pi}}{2},\\
&g_{\Sigma\Sigma\pi}=2\alpha_{NN\pi}g_{NN\pi},\\
&g_{\Sigma\Sigma\eta}=\frac{2}{\sqrt{3}}(1-\alpha_{NN\pi})g_{NN\pi}.
\end{align}
The explicit values are tabulated in Table.~\ref{table0} and they can be determined from $\alpha_{NN\rho}=1.15$~\cite{Ronchen:2012eg}, $\alpha_{NN\pi}=0.4$~\cite{Ronchen:2012eg}, $g_{\rho\rho\rho}=g_{\rho\pi\pi}=6.1994$~\cite{Matsuyama:2006rp,Bando:1987br,Janssen:1994uf},  $g_{NN\rho}=g_{\rho\pi\pi}/2$\cite{Matsuyama:2006rp,Ronchen:2012eg}, $g_{\omega\pi\rho}=11.2$~\cite{Matsuyama:2006rp}, $g_{NN\pi}=1.0$~\cite{Ronchen:2012eg}.  The magnetic couplings $\kappa_{\Sigma\Sigma\omega},\kappa_{\Sigma\Sigma\rho}$, and $\kappa_{\Sigma\Sigma\phi}$ can also be obtained from Ref.~\cite{Ronchen:2012eg} as $\kappa_{\Sigma\Sigma\rho}=\kappa_{\Sigma\Sigma\omega}=1.33$, $\kappa_{\Sigma\Sigma\phi}=-2.35$.

\begin{table}[h!]
\centering
\caption{Coupling constants $g_{\bar{K}^{*}\bar{K}^{*}\alpha}$ and $g_{\Sigma\Sigma{}\alpha}$.}\label{table0}
\begin{tabular}{ccccccc}
\toprule[1.5pt]
                                   &\multicolumn{3}{c}{$V$} &&\multicolumn{2}{c}{$P$} \\
                                   \cline{2-4}               \cline{6-7}
$\alpha$                           &$\rho$ ~~~~~& $\omega$ ~~~~~&$\phi$  &~~~~~&$\pi$  ~~~~~&$\eta$  \\ \hline
$g_{\bar{K}^{*}\bar{K}^{*}\alpha}$ & 3.10  ~~~~~& 3.10     ~~~~~& 4.38  & ~~~~~&5.60   ~~~~~& 9.70     \\
$g_{\Sigma\Sigma{}\alpha}$         & 7.13  ~~~~~& 7.13     ~~~~~& -5.70  &~~~~~&0.80   ~~~~~& 0.69     \\
\bottomrule[1.5pt]
\end{tabular}
\end{table}

With the above Lagrangian terms and coupling constants, the potential for the $\bar{K}^{*}\Sigma$ interaction via the exchange of vector mesons and pseudoscalar mesons are
\begin{align}
i{\cal{V}}_{\lambda^{'}\lambda}^{V}&=f_Ig_{\bar{K}^{*}\bar{K}^{*}V}g_{\Sigma\Sigma{}V}{\cal{H}}^{\mu}_{\lambda_1^{'}\lambda_1}\frac{g_{\mu\nu}-q_{\mu}q_{\nu}/m^2_{V}}{q^2-m^2_{V}}\nonumber\\
                       &\times\bar{u}(k_2^{'},\lambda_2^{'})[\gamma^{\nu}+\frac{\kappa_{\Sigma\Sigma\rho}}{4m_{\Sigma}}(\gamma^{\nu}q\!\!\!/-q\!\!\!/\gamma^{\nu})]u(k_2,\lambda_2),\\
i{\cal{V}}_{\lambda^{'}\lambda}^{P}&=f_I\frac{g_{\bar{K}^{*}\bar{K}^{*}P}g_{\Sigma\Sigma{}P}}{m_{\pi}(q^2-m^2_{P})}\epsilon^{\mu\nu\alpha\beta}k^{'}_{1\mu}\epsilon^{\dagger}_{\nu}(k_1^{'},\lambda_1^{'})k_{1\alpha}\epsilon_{\beta}(k_1,\lambda)\nonumber\\
              &\times\bar{u}(k_2^{'},\lambda_2^{'})\gamma_5q\!\!\!/u(k_2,\lambda_2),
\end{align}
with ${\cal{H}}_{\lambda_1^{'}\lambda_1}^{\mu}=\epsilon'^{\dagger}_1\cdot{q}\epsilon^{\mu}_1+(k_1+k_1^{'})^{\mu}\epsilon'^{\dagger}_1\cdot\epsilon_1-\epsilon'^{\dagger\mu}_1\epsilon_1\cdot{q}-k_1\cdot\epsilon_1
                                        \epsilon'^{\dagger\mu}_1-k'_1\cdot\epsilon'^{\dagger}_1\epsilon^{\mu}_1$.
Here, $u$ and $\epsilon$  are spinor for $\Sigma$ baryon and polarized vector for $\bar{K}^{*}$, respectively.  The $q=k_1^{'}-k_1$,
with $k_{(1,2)}$ and $k^{'}_{(1,2)}$ are the momenta of initial and final $\bar{K}^{*}$($\Sigma$), respectively.  In the center of mass (c.m.) frame, the momenta are
chosen as $k_1=(W-E,0,0,-p)$, $k_2=(E,0,0,p)$ and $k^{'}_1=(W-E^{'},-p^{'}\sin\theta,0,-p^{'}\cos\theta)$, $k^{'}_2=(E^{'},p^{'}\sin\theta,0,p^{'}\cos\theta)$.  The $\theta$
is scattering angle of the outgoing $\bar{K}^{*}$ meson relative to the beam direction.  The $W$ is total energy and $\lambda=\lambda_1-\lambda_1^{'}$ and $\lambda^{'}=\lambda_2-\lambda_2^{'}$, with $\lambda_{(1,2)}$ and $\lambda^{'}_{(1,2)}$ being the helicity of initial and final $\bar{K}^{*}$($\Sigma$), respectively.  For $\rho$ and $\pi$ exchange, the isospin factor $f_{I}$ is $-2$ and $1$ for
isospin $1/2$ and $3/2$, respectively, while the isospin factor $f_I$ is $1$ for all the other exchanges.

The molecular state produced from the $\bar{K}^{*}\Sigma$ interaction corresponds to a pole of the partial wave scattering amplitude ${\cal{M}}$ with fixed spin parity $J^P$.
The scattering amplitude ${\cal{M}}^{J^P}$  in the c.m. frame is calculated using the  one-dimensional Bethe-Salpeter equation~\cite{He:2015mja,He:2014nya,He:2015cca} as
\begin{align}
i{\cal{M}}^{J^P}_{\lambda^{'}\lambda}(p^{'},p)&=i{\cal{V}}^{J^P}_{\lambda^{'}\lambda}(p^{'},p)+\sum_{\lambda^{''}\geq{0}}\int\frac{p^{''2}dp^{''}}{(2\pi)^3}\nonumber\\
                                              &\times{}i{\cal{V}}^{J^P}_{\lambda^{'}\lambda^{''}}(p^{'},p^{''})G_0(p^{''})i{\cal{M}}^{J^{P}}_{\lambda^{''}\lambda}(p^{''},p).\label{eq1}
\end{align}
In order to get the one-dimensional Bethe-Salpeter equation,  the covariant spectator theory is applied by putting the $\Sigma$ on shell~\cite{Gross:2008ps,Gross:2010qm,He:2012zd,He:2011ed}. For details, see Ref.~\cite{He:2015mja}.

The partial wave potential ${\cal{V}}_{\lambda^{'}\lambda}^{J^P}$ can be calculated as
\begin{align}
i{\cal{V}}_{\lambda^{'}\lambda}^{J^P}({\rm p}^{'},{\rm p})&=2\pi\int{}d\cos\theta[d^{J}_{\lambda\lambda^{'}}(\theta)i{\cal{V}}_{\lambda^{'}\lambda}({\bm p}^{'},{\bm p})\nonumber\\
                                              &+\beta{}d^{J}_{-\lambda\lambda^{'}}(\theta)i{\cal{V}}_{\lambda^{'}-\lambda}({\bm p}^{'},{\bm p})].
\end{align}
where  ${\rm p}^{(')}=|{\bm p}^{(')}|$, and the $d^{J}_{\lambda\lambda^{'}}(\theta)$ is the Wigner $d$-matrix. The $\beta=PP_1P_2(-1)^{J-J_1-J_2}$, with $P$ and $P_{1,2}$ being
the parity and $J$ and $J_{1,2}$ being the angular momenta for the system and particle 1 or 2.  In this work we adopt an exponential regularization  in the propagator as $G_0(p)\rightarrow{}G_0(p)[e^{-(k_1^2-m_1^2)^2}/\Lambda^4]^2$ with $k_1$ and $m_1$ being the momentum and the mass of $\bar{K}^{*}$, respectively.
The form factor of  $\Sigma$ is not necessary due to its on-shell-ness.  In our model, there exists only one free parameter, the cutoff $\Lambda$.
And we expect it not far from $1$ GeV, here in a range from 0.6 to 5.0 GeV.

\section{Results and discussions}\label{Sec: results}
Possible bound state from  the $\bar{K}^{*}\Sigma$ interaction can be studied by searching for  poles of the scattering amplitude ${\cal{M}}=V/(1-VG)$.  By analytic continuation
into the complex plane $W\to{}z$, poles will be searched for by varying $z$ in the complex plane to have $|1-V(z)G(z)|=0$~\cite{He:2015mja}.

Since $\bar{K}^*$ and $\Sigma$ carries spin parities $1^-$ and $1/2^+$, respectively, an $S$-wave bound state should have spin parity $1/2^-$  or $3/2^-$. An early experimental analysis ~\cite{Hemingway:1977uw}  suggested  that the $\Xi(2030)$ should carry a spin $J\ge 5/2$. It requires that the $\Xi(2030)$ is at least a $P$-wave  state if
 it is a $\bar{K}^{*}\Sigma$ molecular state.   We would like to note that in our formalism the partial wave decomposition is only on $J^P$ because we work in a relativistic framework. Here we do not distinguish the explicit contributions from $S$ ($P$) and $D$ ($G$) waves.  Generally, the contributions with smaller $L$ is more important for a certain spin parity close to threshold.   In this work, we will consider the spin parities where at least $P$ wave is involved. Since a system with $J\ge 7/2$ will be a $D$- ($G$-) wave state,  only the isospin $1/2$ systems with $J\le 5/2$ will be considered  in this work. Given such constraints,
 possible bound state produced from the $\bar{K}^{*}\Sigma$ interaction are listed in Table.~\ref{table1} with the variation of the cutoff $\Lambda$.

\renewcommand\tabcolsep{0.38cm}
\renewcommand{\arraystretch}{1.5}
\begin{table}[h!]
\centering
\caption{Masses of the bound states with $I=1/2$  from the $\bar{K}^{*}\Sigma$ interaction with different cutoffs $\Lambda$.
Both the cutoff $\Lambda$ and energy $W$ are in units of GeV. }\label{table1}
\begin{tabular}{ccc|ccc}
\toprule[1.5pt]
$J^P$                        &$\Lambda$    & $W$    &$J^P$                        &$\Lambda$     & $W$                        \\ \hline
$\frac{1}{2}^{-}$  & 0.90         &2.084  &$\frac{3}{2}^{+}$  & $1.28$      & 2.073                     \\
                                & 0.95         &2.078  &                                & $1.30$      & 2.063                     \\
                                & 1.00         &2.071  &                                & $1.32$      & 2.051                     \\
                                & 1.10         &2.048  &                                & $1.34$      & 2.036                     \\
\hline
$\frac{3}{2}^{-}$  & $1.00$      & 2.082  &$\frac{5}{2}^{+}$  & $1.24$      & 2.067                        \\
                                & $1.10$      & 2.077  &                                & $1.26$      & 2.058                        \\
                                & $1.20$      & 2.070  &                                & $1.28$      & 2.046                        \\
                                & $1.40$      & 2.050  &                                & $1.30$      & 2.031                         \\
\bottomrule[1.5pt]
\end{tabular}
\end{table}

In the six cases considered, four bound states  with isospin $1/2$ can be found in the $\bar{K}^*\Sigma$ system with spin parities $1/2^-$, $3/2^-$, $3/2^+$ and $5/2^+$.  As in other systems~\cite{He:2015mja,He:2015cea,He:2016pfa,He:2015cca} , the binding energies of the bound states  produced from the  $\bar{K}^*\Sigma$ interaction increase with the cutoff $\Lambda$.  Among these four bound states, the $1/2(1/2)^-$  and $1/2(3/2^-)$ states are from $S$ and/or $D$ wave while the $1/2(3/2^+)$ and $1/2(5/2^+)$ states from $P$ and/or $G$ wave. Since the contributions with smaller $L$ is more important for a certain spin parity close to threshold,
one can say that there exist two $P$-wave bound states and two S-wave bound states from the $\bar{K}^*\Sigma$ interaction.  The two negative-parity states in $S$ wave  appear with cutoffs about 1 GeV and two positive-parity states in $P$ wave appear with cutoffs about 1.2 GeV.  It shows that near threshold a larger cutoff is needed to enhance the attraction  for the $P$-wave system than for the $S$-wave system. Such a phenomenon is also found in the case of the $\bar{D}^*\Sigma$ interaction~\cite{He:2016pfa}.

\begin{figure}[h!]
\begin{center}
\includegraphics[scale=0.65]{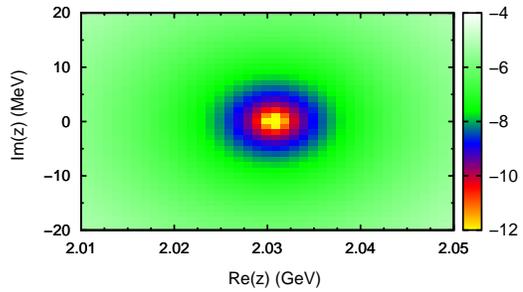}
\caption{The pole position of the scattering amplitude for the quantum number $I(J^P)=1/2(5/2)^{+}$ as a function of $z$.}
\label{t-mass}
\end{center}
\end{figure}

The experimental analysis~\cite{Hemingway:1977uw} favors a spin not less than $5/2$ for the $\Xi(2030)$, which is also suggested by PDG~\cite{Patrignani:2016xqp}.  As discussed above, the system with a spin parity $5/2^-$ and spin $J\ge7/2$ is at least a $D$-wave state, whose contribution should be small.  The study in Ref.~\cite{He:2016pfa} suggested that $P$-wave contributions are still considerable and may be observed in experiment. The only possible spin parity to interpret the $\Xi(2030)$ in our molecular  picture is $5/2^+$.    Fortunately,  a bound state can be found at a cutoff about 1.3 GeV in our model as expected, and with the increase of the cutoff its
mass can reach 2030 MeV.   The pole with $I(J^P)=1/2(5/2)^{+}$  is illustrated in Fig.~\ref{t-mass}.

For the $\Xi(2120)$, the experimental information is scarce, which makes its assignment to a certain state difficult.  In our study, two states with $1/2(1/2^-)$ and $1/2(3/2^-)$ are produced from the $\bar{K}^*\Sigma$ interactions,  which is consistent with the results of the chiral unitary approach~\cite{Oset:2009vf,Gamermann:2011mq,Khemchandani:2016ftn}. In Ref.~\cite{Khemchandani:2016ftn}  a  bound state with $3/2^-$ arises at $2046-i8.2$,which is consistent with our results. In addition to these two $S$-wave negative-parity states, a $P$-wave positive-parity state with $1/2(3/2^+)$ is also produced from the $\bar{K}^*\Sigma$ interaction.  Because of the lack of  experimental information
on the spin parity of the $\Xi(2120)$, any of the three bound states can be a candidate for the $\Xi(2120)$.

Above, we presented the results  for the isospin 1/2 channel, which corresponds to the $\Xi$ baryon. In the $qqq$ picture, the isospin of a $\Xi$ baryon is 1/2 because there is only one $u/d$ quark. However, in the $\bar{K}^*\Sigma$ system, there exist two $u/d$ quarks and a $\bar{u}/\bar{d}$ antiquark. Hence, an isospin 3/2 is allowed for a molecular state, which is analogous to the existence of a charged charmonium-like state. In Table~\ref{table3}, we list the bound states with isospin 3/2 from the $\bar{K}^*\Sigma$ interaction.  Only two bound states with $J^P=1/2^-$ and $J^P=1/2^+$ can be produced with cutoffs about 1 GeV.

\begin{table}[h!]
\centering
\caption{Masses of the bound states with $I=3/2$  from the $\bar{K}^{*}\Sigma$ interaction with different cutoffs $\Lambda$.
Both the cutoff $\Lambda$ and energy $W$ are in units of GeV. }\label{table3}
\begin{tabular}{ccc|ccc}
\toprule[1.5pt]
$J^P$                        &$\Lambda$    & $W$    &$J^P$                        &$\Lambda$     & $W$                        \\ \hline
$\frac{1}{2}^{-}$  & $1.10$      & 2.082                 &$\frac{1}{2}^{+} $  &0.94          & 2.082                      \\
                                &$1.15$      & 2.073     &                                &0.97          & 2.071                      \\
                                & $1.20$      & 2.058    &                                &1.00          & 2.056                      \\
                                & $1.25$      & 2.038    &                                &1.03          & 2.036                      \\
\bottomrule[1.5pt]
\end{tabular}
\end{table}

\section{ Summary}
Inspired by the LHCb observation of the pentaquark states and their molecular interpretations,
we have studied possible bound states from the $\bar{K}^{*}\Sigma$ interaction by solving a quasipotential Bethe-Salpeter equation  within the one-boson exchange model.  A bound state with the quantum number $I(J^{p})=1/2(5/2^{+})$ from the  $\bar{K}^{*}\Sigma$ interaction
is produced at $\Lambda=1.3$ GeV.  This bound state can be associated to the $\Xi(2030)$ as a $P$-wave molecular state.   Three other   bound states with quantum numbers $J^{P}=1/2(1/2^{-})$, $1/2(3/2)^-$ and $1/2(3/2^+)$ are also produced from the $\bar{K}^{*}\Sigma$ interaction.

For the $\Xi(2030)$,  our study showed that it could be  a $P$-wave $\bar{K}^*\Sigma$ bound state with spin parity $5/2^+$, consistent with the available experimental information.
On the other hand, we do not have enough experimental information to determine which bound state produced from the $\bar{K}^*\Sigma$ interaction is the $\Xi(2120)$. If we follow the assignment in Ref.~\cite{Khemchandani:2016ftn}, that the $\Xi(2120)$ is a  molecular bound state with $1/2(3/2^-)$, it is interesting to see that the $\Xi(2120)$ and $\Xi(2030)$ exhibit a pattern analogous to the $P_c(4380)$ and $P_c(4450)$ in Ref.~\cite{He:2016pfa}, where it was found that the $P_c(4450)$ and $P_c(4380)$ can be produced from the $\bar{D}^*\Sigma_c$ interaction as a $P$-wave and $S$-wave state with $5/2^+$ and $3/2^-$, respectively.

Clearly,  more experimental efforts are needed to better understand the nature of the $\Xi(2120)$ and test  the scenario proposed in the present study. There exist plans to study  double-strangeness baryons at facilities such as JLab, J-PARC, and \={P}ANDA.  We strongly recommend our experimental colleagues to study  the $\Xi(2030)$, $\Xi(2120)$ and the double-strangeness baryons with isospin $3/2$. These studies can also help improve our understand of the LHCb pentaquark states.

\section{Acknowledgments}
This work is partly supported by the National Natural
Science Foundation of China under Grants No.11522539, 11735003,
11675228,11475227 and the fundamental Research Funds for
the Central Universities.

\end{document}